\documentclass[aps,twocolumn,prl,superscriptaddress,showpacs]{revtex4}

\usepackage{latexsym,amsmath,verbatim,amsfonts}
\usepackage{graphicx}

\newcommand{\av}[1]{\langle {#1} \rangle}

\begin{document} 

\title{Modeling temporal networks using random itineraries}

\author{Alain Barrat}
\affiliation{Centre de Physique Th\'eorique, Aix-Marseille Univ, CNRS
  UMR 7332, Universit\'e de Toulon, 13288 Marseille cedex 9, France}
\affiliation{Data Science Lab,
Institute for Scientific Interchange (ISI) Foundation, Torino, Italy}

\author{Bastien Fernandez}
\affiliation{Centre de Physique Th\'eorique, Aix-Marseille Univ, CNRS
  UMR 7332, Universit\'e de Toulon, 13288 Marseille cedex 9, France}

\author{Kevin K.~Lin}
\affiliation{Department of Mathematics and Program in Applied
  Mathematics, the University of Arizona, Tucson, AZ 85721, USA}

\author{Lai-Sang Young}
\affiliation{Courant Institute of Mathematical Sciences, New York University,
New York, NY 10012, USA}

\begin{abstract}
We propose a procedure to generate dynamical networks with bursty,
possibly repetitive and correlated temporal behaviors. Regarding any
weighted directed graph as being composed of the accumulation of
paths between its nodes, our construction uses random walks
of variable length to produce time-extended structures with adjustable
features. The procedure is first described in a general framework. It
is then illustrated in a case study inspired by a transportation
system for which the resulting synthetic network is shown to
accurately mimic the empirical phenomenology.
\end{abstract}

\pacs{89.75.-k, 89.75.Hc, 02.50.Ey , 05.40.Fb , 89.75.Fb}

\maketitle 

Many systems in nature or related to human activities are conveniently
represented as networks of interacting units
that can exchange material or information. 
This approach, combined with techniques from graph
theory, statistical physics and data analysis, has led to countless
interesting studies and insights in numerous fields
\cite{science,Dorogovtsev:2003,Newman:2003,Pastor:2004,Caldarelli:2007,Barrat:2008}.

Until recently, network structures were often regarded as being
stationary, both for simplicity and because of scarceness
of datasets on the time variability of connectivity patterns. Thanks
to advanced acquisition technologies and large scale production of
time-resolved data, temporal information has become more accessible
in numerous contexts, from communication networks
\cite{Eckmann:2004,Onnela:2007,Makse:2009,Amaral:2009,Karsai:2012} to
proximity patterns \cite{Cattuto:2010,Salathe:2010} and infrastructure
networks \cite{Gautreau:2009,Bajardi:2011}.
This has led to the recent 
surge of activity in the field of ``temporal networks'' \cite{review_holme}.
Data analysis revealed the coexistence of statistically stationary
properties and local variations, interaction burstiness 
\cite{Eckmann:2004,Holme:2005,Onnela:2007,Makse:2009,Amaral:2009,Gautreau:2009,Cattuto:2010,Salathe:2010,Bajardi:2011,Karsai:2012,humdyn-barabasi,humdyn-vazquez,Barabasi:2010,review_holme}
and the occurrence of non-local repetitive 
patterns \cite{Bajardi:2011,kovanen:2011}. These structural properties affect the dynamical processes taking place on
networks \cite{Vazquez:2007,Iribarren:2009,Miritello:2011,Isella:2011,Karsai:2011,Bajardi:2012,Holme:2013,Hoffmann:2012,Perra:2012PRL,Ribeiro:2012}. Therefore, 
they have to be accounted for in modeling approaches.

Despite the proliferation of temporally resolved datasets, strong
limitations perdure. Indeed, data are often only accessible in 
restricted forms, such as single samples of limited sizes and statistical relevance.
Comparison of connection patterns at different times is not always possible.
Some datasets consist only
of aggregated information and do not provide access to the temporal course of events. In such circumstances, it is necessary to be able to generate synthetic time-extended structures whose aggregation would reproduce the data at hand. This would enable one to go beyond the approximation of 
static networks and to incorporate dynamical components in network structures. 

Several models of temporal networks have been proposed in
the literature \cite{GB08,Gautreau:2009,Latora:2010,Stehle:2010,Perra:2012,Starnini:2013}. Their dynamics mostly consist of  link updates and show
 that a global complex space-time organization can emerge as the result of simple
pairwise cooperation or competition rules between individual units. 
Dataset randomization is also employed to create null models against which the
temporal pattern complexity of empirical data can be evaluated
 \cite{review_holme,Karsai:2011}. 

The modeling approach developed in this Letter is to some extent more
direct and pragmatic. It is intended to provide a versatile procedure for
the construction of time-dependent networks with adjustable
characteristics, that mimic bursty and possible repetitive behaviors, as well as extended temporal correlations, as they are observed in real world systems. The goal is to obtain realistic temporal
structures independently of any access to datasets or assumptions about basic
interaction mechanisms.

\begin{figure}[t]
\begin{center}
\includegraphics[scale=0.23]{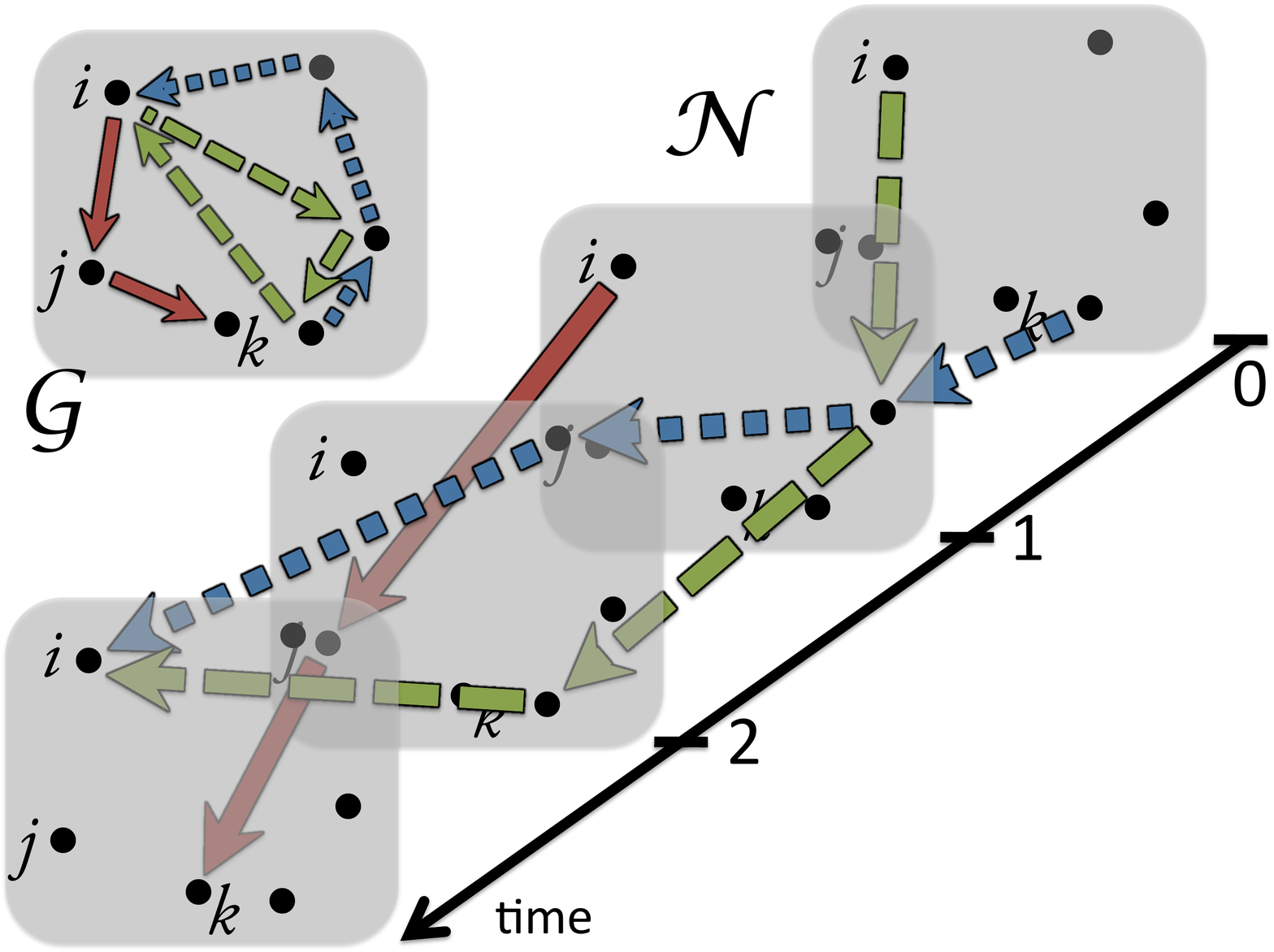}
\caption{(color online).
Schematic representation of the construction procedure: a weighted directed graph ${\cal G}$
can be regarded as a superposition of paths. Unfolding these paths in time results in itineraries (of variable characteristics) that generate a temporal network ${\cal N}$.}
\label{fig:stnetwork}
\end{center}
\end{figure}

The proposed procedure originates from the observation that in many 
graphs representing systems such as transportation, trade or 
communication networks, connection patterns are the result of activity 
spreading along noisy itineraries within a complex structure.  
The weighted directed graph used to represent this activity is then given by
the accumulation of such itinerary traces.  Our
construction proceeds in the opposite manner. Starting from a given weighted directed
graph ${\cal G}$, it assumes that this graph is the superposition of 
random walks in  a given time window ${\cal W}$,
and  seeks to ``unfold'' these walks,
as schematically illustrated in Fig.~\ref{fig:stnetwork}.

Below, we first describe the construction in a general setting.  We then apply it
to a case study inspired by a 
transportation system,
and use this real-world system to benchmark the synthetic network. Finally, we put forward a characterization of the topological and temporal correlations that takes into account
the heterogeneity of the activity in the network.

\paragraph{Temporal network construction.}
The construction takes as input a graph ${\cal G}=({\cal S},{\cal E})$ and a time window ${\cal W}=\{0,1,\cdots ,T-1\}$. Here, ${\cal S}$ is a set of nodes and ${\cal E}$ is a set of weighted
directed edges, either obtained from data or built with prescribed
properties such as  in- and out-degree distributions and/or distributions of link weights (for definitions of degree, weight and strength, 
see {\sl e.g.}\ \cite{Barrat:2008}). The temporal network is built as follows.

(i)  First, the following random walk characteristics are chosen:
distributions of starting times in ${\cal W}$, of starting locations in
${\cal S}$, of walk lengths, and of
residence times at each node.

(ii) Then, the random walks are generated independently, one after another, using the characteristics specified above. 
Each walk defines an itinerary that consists of a list of events 
$\{(i_0,i_1,t_1),(i_1,i_2,t_2),\cdots,(i_{\ell-1},i_\ell,t_\ell)\}$ 
with $t_1 < \cdots < t_\ell$  (Fig.~\ref{fig:stnetwork}). At each step,
the node $i_p$ is uniformly chosen among the out-neighbors of $i_{p-1}$,
and the residence time $t_{p+1} - t_{p}$ is drawn from the residence
  time distribution associated with $i_{p}$. 
Any walk reaching a node with no out-link terminates there, even if it has not yet reached its prescribed length.

(iii) As the construction proceeds, 
the graph ${\cal G}$ is continuously altered as follows: if a walk passes through the link $i\to j$, the weight $w_{ij}$ is 
decreased by 1; 
whenever $w_{ij}$ reaches $0$, the edge $i\to j$ is discarded from the graph and 
cannot be used anymore. 

(iv) Walks are generated until all edges have been discarded from ${\cal G}$. Then, the process terminates.

The temporal network ${\cal N}$ is defined as the union of all the constructed itineraries.
By construction, 
the set of weighted directed edges resulting from the collection of all events in ${\cal N}$ 
coincides with the original set ${\cal E}$. Itineraries are interpreted differently depending upon context. 
For instance, in transportation systems, the
event $(i,j,t)$ is regarded as a material displacement from $i$ to $j$
at time $t$. In communication networks, events
correspond to transmission of information.

Note also that the construction can be used to model routine/seasonal processes occurring in 
consecutive time windows,  using a ``noisy deterministic''
rule: for each $n$, a tunable fraction of walks is redrawn in
${\cal W}_n=\{nT,nT+1,\cdots, (n+1)T-1\}$, while the remaining itineraries are repeated
identically from ${\cal W}_{n-1}$ to ${\cal W}_{n}$.

We claim that, under suitable choices of input parameters, the network ${\cal N}$ 
can mimic real-world features, in particular bursty temporal patterns and extended correlations. 
To support these assertions, we benchmark the synthesized network in a
case study inspired by a cattle trade system analysis
\cite{Bajardi:2011}; note this example is aimed to illustrate the construction algorithm, rather than to accurately fit specific datasets.

\paragraph{Case study.}  
In this application, we consider $|{\cal S}|=10^4$ nodes representing farms. The graph ${\cal G}$ is
constructed using a variant of the uncorrelated configuration model
\cite{Catanzaro:2005}. Following statistics reported in
\cite{Bajardi:2011}, we use as in- and out-degree distributions the
power-law $P(k_{in,out})\propto k_{in,out}^{-2.5}$ (with cut-off at
$\sqrt{|{\cal S}|}$ to avoid degree correlations
\cite{Catanzaro:2005}).  Weights are also power-law distributed
according to $P(w) \propto w^{-2.5}$. Interpreting events in ${\cal
  N}$ as material displacements, we impose flux conservation at
every node so that in- and out-strengths
balance, {\sl i.e.}\ $\sum_jw_{ji}=\sum_jw_{ij}$ holds for almost all
$i\in {\cal S}$ (in
practice, more than $99\%$ of ${\cal S}$. 

\begin{figure}[t]
\begin{center}   
  \begin{scriptsize}
    \resizebox{2.7in}{!}{\includegraphics*[viewport=0.1in 0in 4.4in 1.5in]{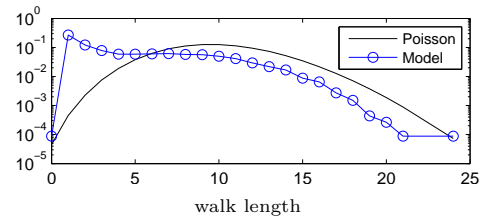}}\\
    walk length
  \end{scriptsize}
\caption{(color online). Prescribed (solid line) and realized
  (circles) distributions of random walk lengths in the temporal
  network ${\cal N}$. The discrepancy is due to the decay of the
  number of remaining edges in ${\cal G}$ as the construction
  proceeds.}
\label{DistribRW}
\end{center}
\end{figure}

We consider a temporal window ${\cal W}$ of length $T=10^3$ units (``days''), with periodic boundary conditions, to generate
${\cal N}$ from ${\cal G}$. To this aim, we use random walks with uniformly distributed
starting times and locations.
We assign each node $i$ a residence time $\tau_i$, drawn from
  $P(\tau_{i})\propto \tau_{i}^{-3}$ for $1 \le \tau_{i} \le 60$; a walk
  visiting node $i$ stays for time $\tau_{i}$, plus an additional random
  delay drawn from a Poisson distribution with mean $\tau_i/5$.  (The
  power-law $P(\tau_i)$ is inspired by \cite{Bajardi:2011}; similar
  results are obtained for an exponential $P(\tau_{i})$.)  Walk lengths are generated using a Poisson
distribution with average $10$. Notice that since the number of graph
edges decreases as the construction proceeds, long walks can seldom be
achieved when few edges remain and the realized distribution accordingly
exhibits an excess of short paths, see Fig.~\ref{DistribRW}.

Three main types of analysis have been proposed
 in order to study structural and temporal properties of a time-resolved transportation network \cite{Gautreau:2009,Bajardi:2011,kovanen:2011}: 
(i) statistics of aggregated networks on various time scales,
(ii) distributions of activity and inactivity periods and (iii) repetition of
connection patterns such as temporal paths. Here, we show that when applied to the synthetic 
network ${\cal N}$, these diagnostics proffer characteristics very similar to those in
empirical data. 
\begin{figure}[ht]
\begin{center}
  \begin{scriptsize}
    \begin{tabular}{ccc}
      \resizebox{1.1in}{2.2in}{\includegraphics[viewport=0.15in 0in 2.35in 5in]{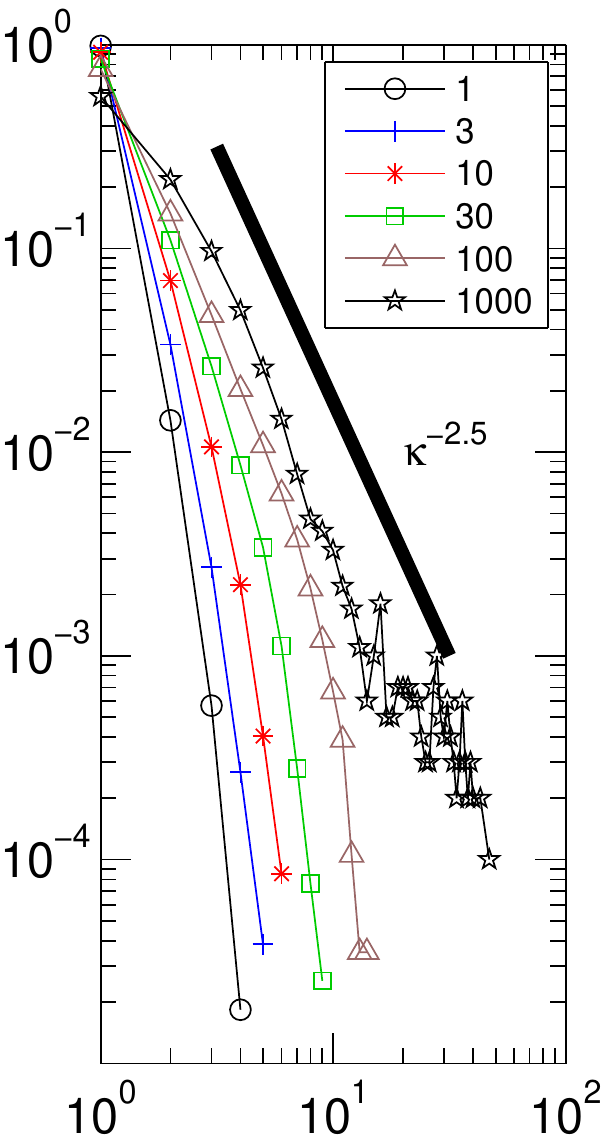}}&
      \resizebox{1.1in}{2.2in}{\includegraphics[viewport=0.15in 0in 2.35in 5in]{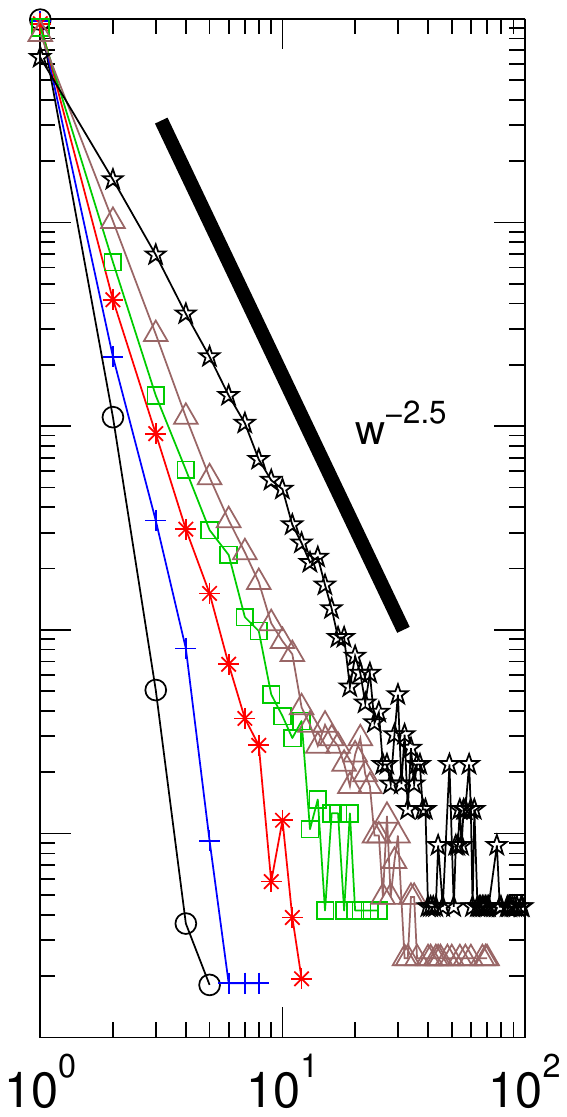}}&
      \resizebox{1.1in}{2.2in}{\includegraphics[viewport=0.15in 0in 2.35in 5in]{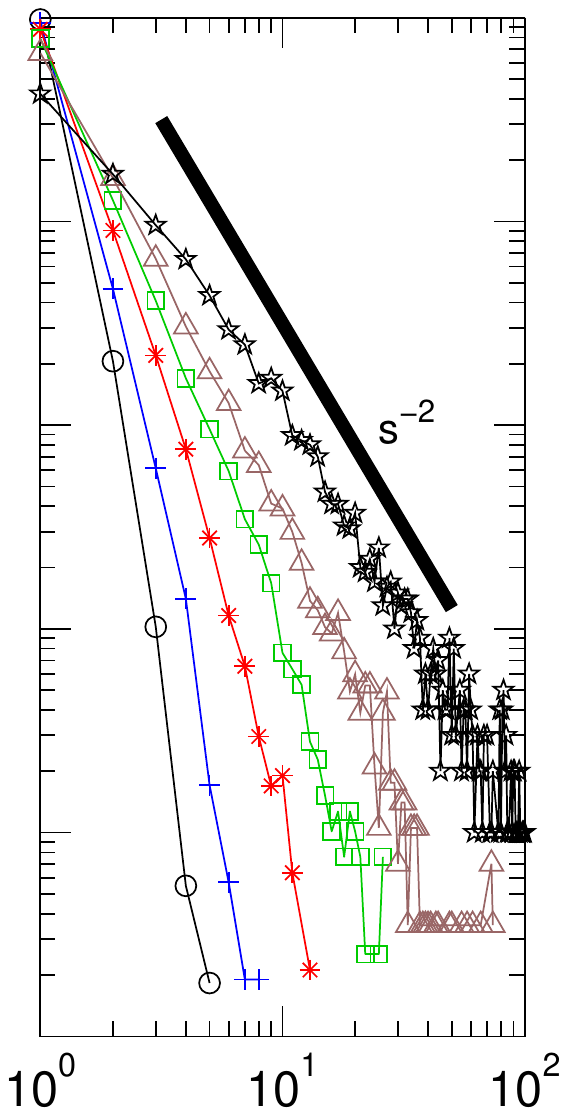}}\\[-3ex]
      in-degree $\kappa$ &
      weight $w$ &
      strength $s$\\
    \end{tabular}
  \end{scriptsize}
\caption{(color online). Distributions of node in-degrees, link
  weights, and node strengths for networks aggregated over intervals
  of various lengths. For integration lengths 
  $<10^3$, the figures display 
  distributions obtained by averaging over all corresponding intervals in ${\cal W}$.  Power
  laws (thick black lines) with expected exponents are plotted for
  reference.}
\label{STATICDISTRIB}
\end{center}
\end{figure}

Real world data exhibit robust heterogeneous behaviors at all temporal scales, 
as asserted by power-law distributions of in-degrees, strengths and link weights \cite{Gautreau:2009,Cattuto:2010,Bajardi:2011}. 
These distributions remain almost stationary when integrated over distinct windows of equal length. 
As shown in Fig.~\ref{STATICDISTRIB}, the same properties are observed in ${\cal N}$.
These plots also validate the construction algorithm: statistics aggregated over the whole ${\cal W}$
reveal power-law distributions with slopes in close agreement with the corresponding ones in ${\cal G}$. 
\begin{figure}[ht]
\begin{center}   
  \begin{scriptsize}
    \begin{tabular}{cc}
      \resizebox{1.675in}{1.5in}{\includegraphics[viewport=0.2in 0in 3.8in 4in]{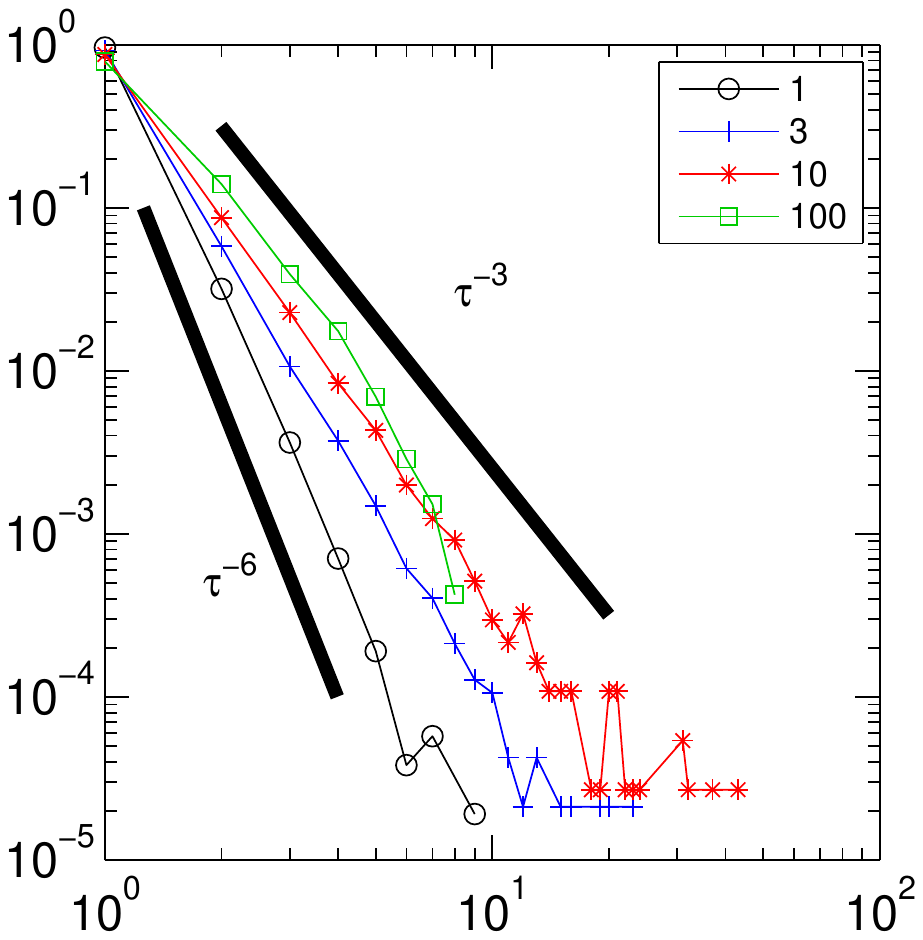}}&
      \resizebox{1.675in}{1.5in}{\includegraphics[viewport=0.2in 0in 3.8in 4in]{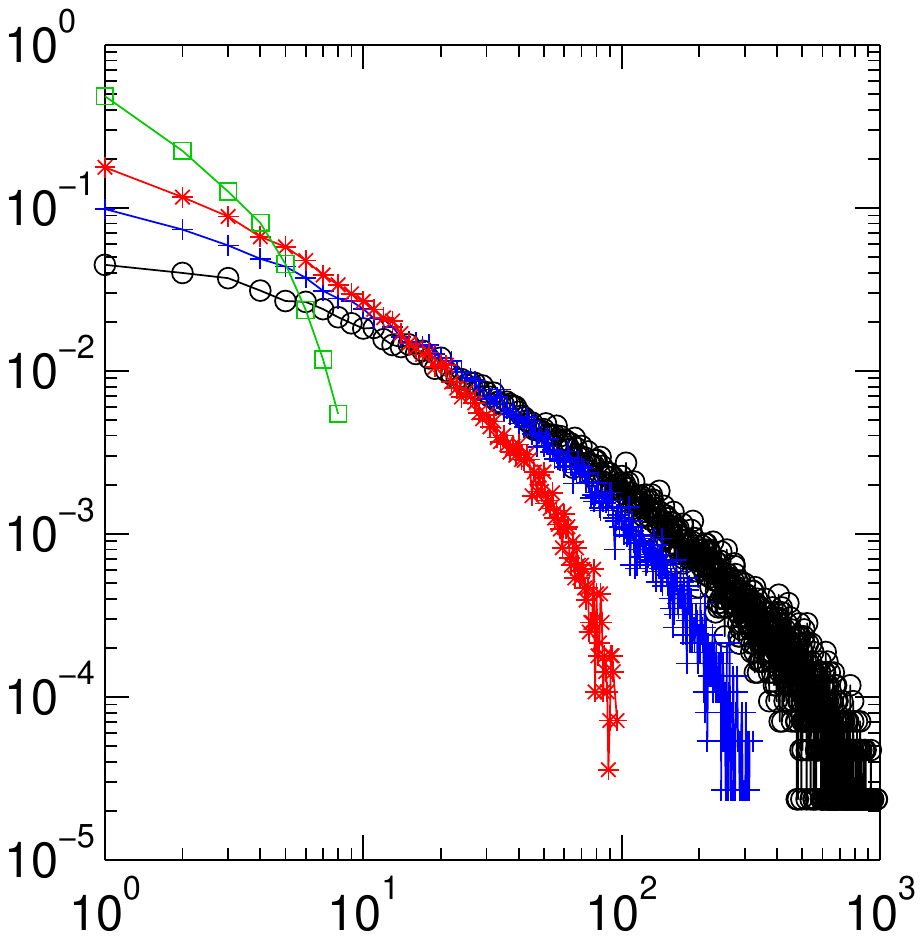}}\\[-2.5ex]
      $n$ & $n$ \\
      (a) Node activity time distrib.\ &
      (b) Node inactivity time distrib.\ \\
    \end{tabular}
  \end{scriptsize}
\end{center}
\caption{(color online).  Distributions of activity (a) and inactivity (b) periods aggregated over 
intervals of various lengths $T_\text{agg}=1, 3, 10, 100$, {\sl i.e.}\  fractions of nodes that are active 
at least once in each of (resp.\ inactive during) $n$ consecutive intervals of duration $T_\text{agg}$
(the maximal value of $n$ is $T/T_\text{agg}$).}
\label{ACTIV-INACTIVDISTRIB}
\end{figure}

Robust heterogeneous features in transportation networks have also
been observed in activity and inactivity period statistics, {\sl
  i.e.}\ in the distributions of the periods during which a node (or a link) 
is continuously active or inactive. 
Such observables typically exhibit broad distributions 
\cite{Gautreau:2009}. Fig.~\ref{ACTIV-INACTIVDISTRIB}
reports these distributions for nodes, obtained from ${\cal N}$ using
various aggregation scales $T_\text{agg}$, from the most detailed (``daily'', $T_\text{agg}=1$)
resolution to coarser timescales mimicking weekly or monthly aggregations (similar
results are obtained for  links).
Activity distributions reveal power-law behaviors in quantitative agreement with empirical data \cite{Bajardi:2011},
with decay exponents in the same range of values 
(increasing from $-6$ to $-3$ as the aggregation interval
$T_\text{agg}$ increases).
Moreover, distributions of inactivity periods are broader, more concave, 
and extend across all timescales,
as in  \cite{Bajardi:2011}.

\begin{figure}[t]
\begin{center}
\begin{tabular}{cc}
\resizebox{1.7in}{!}{\includegraphics[viewport=0.1in 0in 2.9in 2in]{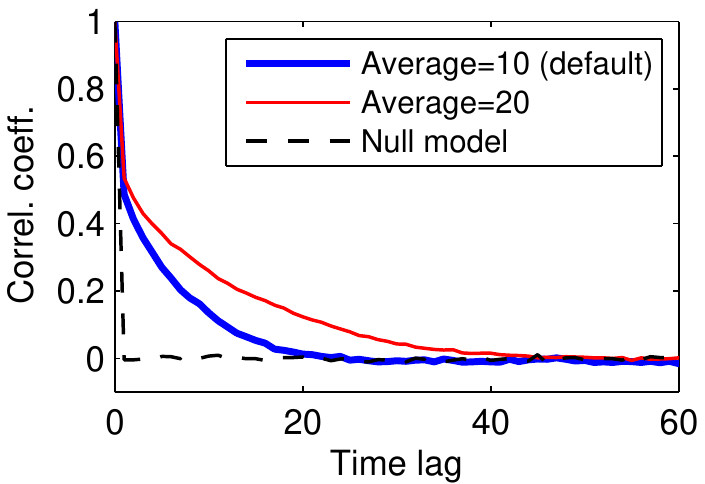}}&
\resizebox{1.7in}{!}{\includegraphics[viewport=0.1in 0in 2.9in 2in]{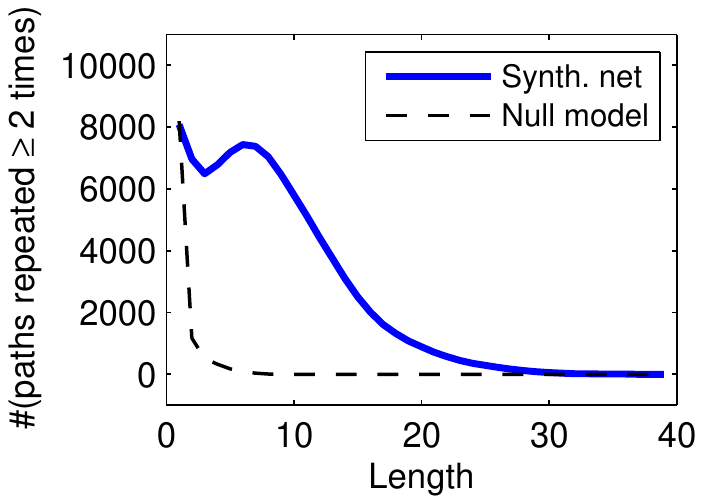}}
\end{tabular}
\caption{(color online). {\sl Left.}\ Autocorrelation function $C(t)$
  {\sl vs.}\ $t$, averaged over $40$ realizations of ${\cal N}$, for
  two distinct walk length averages
  and compared to a null model where all links have been temporally randomized,
  equivalent to imposing that all walks in the construction have length 1. {\sl Right.}\ Number of $\tau$-tolerant temporal paths
  ($\tau=5$) occurring at least twice {\sl vs.}\ their length, both in
  ${\cal N}$ and in the null model.}
\label{fig:paths}
\end{center}
\end{figure}

In addition to burstiness, the activity in ${\cal N}$ shows positive temporal
correlations, as can be anticipated from the
nature of the construction. In particular, global activity
correlations can be appreciated by using 
\[
C(t) = \frac{\av{A(\cdot)A(\cdot+t)}_\text{ti}-\av{A(\cdot)}_\text{ti}^2}{\text{Var}(A(\cdot))}
\]
where $A(t)$ denotes the total number of events $(i,j,t)$ on day $t$ and
$\av{\cdot}_\text{ti}$ means time averaging over ${\cal W}$. As  
 Fig.~\ref{fig:paths} (left) shows, $C(t)$ is positive for all
  $t$ in an extended interval, 
  whereas its analog in a null model obtained
  by reshuffling all links  is 0 for any $t\neq0$. Note that such a null model is equivalent to 
  a temporal network construction in which we impose all the random walks to have length $1$.

Beside global correlations, local structures can be
highlighted by exhibiting chronological link sequences
that occur more often than by chance
\cite{Bajardi:2011,kovanen:2011}. A typical example is given by
$\tau$-tolerant temporal paths, {\sl i.e.}\ paths composed of
consecutive active links within $\tau$ days of each other.
The  number of such paths is much larger in empirical temporal networks than in 
reshuffled data \cite{Bajardi:2011}. (Note that only the case $\tau=1$ was considered in 
\cite{Bajardi:2011}.) Fig.~\ref{fig:paths}
(right), which shows the number of $\tau$-tolerant temporal paths that
occur at least twice in ${\cal N}$, possesses similar features, 
providing further evidence of the presence of strong
correlations in the synthetic network.

\paragraph{Detailed activity correlations} 
The nature of the temporal network construction, which relies on random walks, suggests that non-trivial local activity correlations should also simultaneously emerge at both topological and temporal levels. Indeed, if a node is active at a certain time, then an elevation of activity should be detected in its topological and temporal neighborhood. (By contrast, such correlations are nonexistent in a model where all links have been reshuffled.) Here, we provide quantitative estimates of  these correlations.  

In heterogeneous networks such as the current synthetic ${\cal N}$, local activity patterns are extremely diverse and are hardly captured by global averages such as $C(t)$. Therefore, prior to averaging, we first need to divide the nodes of ${\cal S}$ into categories according to their total activity (defined as the total number of events in ${\cal W}$, starting from the considered node).
The activity at a node in ${\cal S}$ ranges from $1$ to over $200$, with average $|{\cal N}|/|{\cal S}|\sim 6$. 
Accordingly, we define a node as ``busy'' if its integrated activity exceeds a given threshold, say {\sl e.g.}\ $40$ (results are qualitatively insensitive to this choice). 
Furthermore, non-busy nodes are categorized into level sets of the distance $d$ to the set of busy nodes 
${\cal B}$, where $d$ is measured following the edges in ${\cal E}$. 
Altogether,
the categories 
${\cal B}=\{d=0\}$, $\{d=1\}, \ \dots, \ \{d=4\}$ account for more than $99\%$ of $\cal S$.

For each node category ${\cal C}$, the correlations between the activity in nodes of ${\cal C}$ at time $t$ and in their neighborhood 
at time $t+\tau$ are then measured by 
\[
F_{{\cal C},r}(\tau)=\av{a_{B(i,r)}(t+\tau)}_{\{ i \in {\cal C}, t\in {\cal W} | s_\text{out}(i,t)>0 \}}
\] 
where $a_{E}(t)= \sum\limits_{j \in E}s_\text{out}(j,t) / |E|$ is the average activity per node 
in a subset of nodes $E$ at time $t$ ($s_\text{out}(j,t)$ is the number of events starting from $j$ at time $t$), $B(i,r)$ is the ball of center $i$ and radius $r$, and 
the average $\av{\cdot}$ is taken over all $i \in {\cal C}$ and $t
\in {\cal W}$ such that $s_\text{out}(i,t)>0$.

\begin{figure}
\begin{center}
  \begin{tiny}
    \begin{tabular}{cc}
      ${\cal C}={\cal B}, r=1$ & ${\cal C}={\cal B}, r=2$\\[-2ex]
      \resizebox{1.5in}{!}{\includegraphics[viewport=0in 0in 3in 2in]{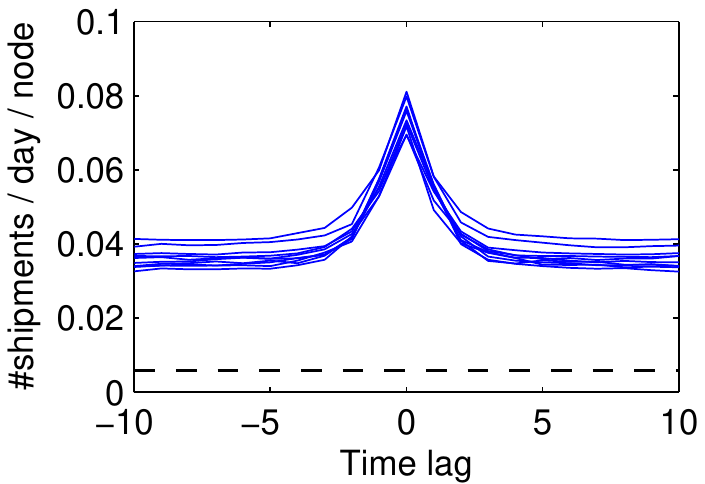}}&
      \resizebox{1.5in}{!}{\includegraphics[viewport=0in 0in 3in 2in]{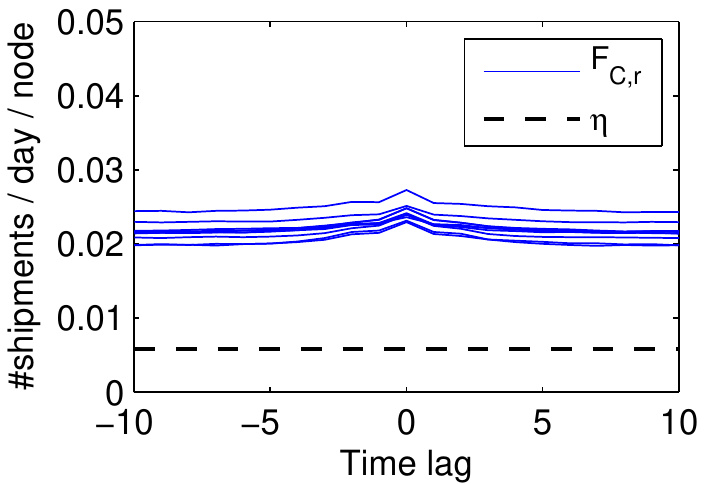}}\\[1ex]
      ${\cal C}=\{d=2\}, r=1$ & ${\cal C}=\{d=2\}, r=2$\\[-2ex]
      \resizebox{1.5in}{!}{\includegraphics[viewport=0in 0in 3in 2in]{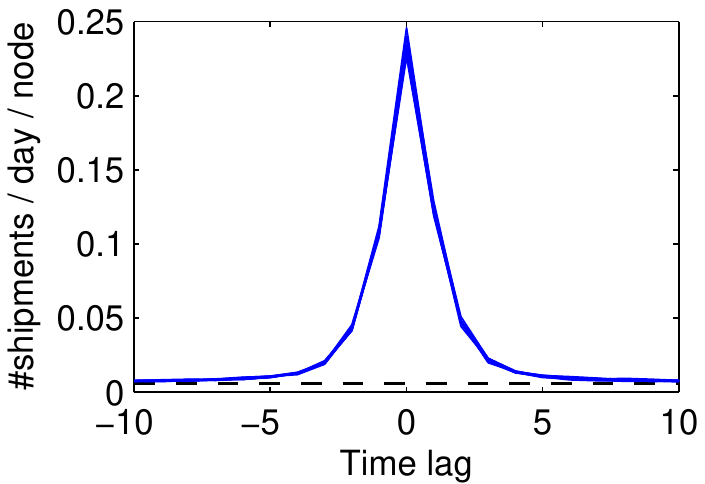}}&
      \resizebox{1.5in}{!}{\includegraphics[viewport=0in 0in 3in 2in]{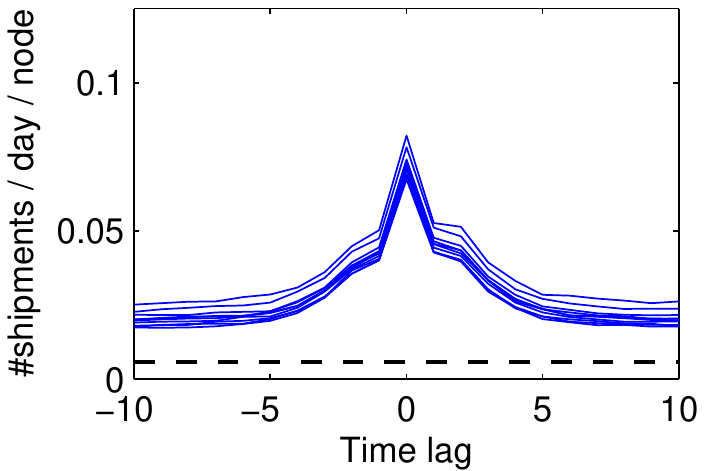}}\\
    \end{tabular}
  \end{tiny}
\end{center}
\caption{(color online). $F_{{\cal C},r}(\tau)$ {\sl vs.}\ $\tau$ for 2 categories of nodes ${\cal C}={\cal B}$ (busy nodes) and $\{d=2\}$, and two neighboring ball radii $r=1$ and $r=2$.  Each panel
  displays the curves corresponding to $10$ realizations of ${\cal N}$.  Dashed lines refer to the average
  activity $\eta=|{\cal N}|/T/|{\cal S}|$ per day per node.}
\label{fig:space-time}
\end{figure}

Fig.~\ref{fig:space-time} displays ``activity spike curves''
$F_{{\cal C},r}(\tau)$ {\sl vs.}\ $\tau$ for two categories, ${\cal B}$ and
$\{d=2\}$, and $r=1,2$. In a null model where all links are reshuffled (equivalent to a model
built with random walks of length $1$), each such curve would 
be perfectly flat except at $\tau=0$, where it would take a larger value. In contrast,
each curve has a distinctive shape here that is almost
insensitive to the network realization. Hence, for each category, these curves provide 
a characterization of activity correlation patterns in a topological and temporal neighborhood.
 
The peaked shape of $F_{{\cal C},r}(\tau)$ illustrates how the 
occurrence of activity in the center of a ball is accompanied by an elevated 
activity in the ball for a number of days before and after. The peaks are substantially sharper for smaller balls (of radius 1), and are instead barely detectable when averaging in the ball of radius 2 around ${\cal B}$.  
At large values of $\tau$, $F_{{\cal C},r}(\tau)$ approaches a well-defined baseline level of activity. 
For $r=1$, this baseline is higher around busy nodes
than around $\{d=2\}$, pointing to a higher level of mean activity in the neighborhood of ${\cal  B}$, as it can be anticipated.  

\paragraph{Concluding remarks.} In this Letter, a simple framework for the construction of temporal networks displaying bursty, repetitive and correlated behaviors has been presented, based on random walks on a predefined aggregated graph. The construction is sufficiently versatile to adapt to a wide range of situations, by appropriate tuning of input parameters. In particular, it would be interesting to apply it to initial graphs ${\cal G}$ with community structure. As random walks tend to be trapped inside communities, the construction is expected to naturally give rise to the emergence of temporal communities. Another possible extension would be to use, instead of random walks as building blocks, spreading trees of propagation processes on networks.  

Our procedure enables one to obtain plausible and realistic instances of temporal networks when only the aggregated structure is known. Hence, it can be employed to compensate for the lack of time-resolved data or to provide alternative scenarios
when access to empirical information is limited. Moreover, that the synthetic network has tunable properties is essential to assessing the influence of time-dependent features on the dynamical processes on networks, especially when only aggregated information
is available. Finally, structural flexibility also makes it possible to test the relevance of
various definitions of centrality measures
for nodes and links in temporal networks.

\textit{Acknowledgments.} AB and BF are partly supported by FET project
MULTIPLEX 317532 and by CNRS PEPS {\em Physique Th\'eorique et ses
  Interfaces}.  LSY is supported in part by NSF grant DMS-1101594, and
KL by NSF grant DMS-0907927.

\end{document}